\documentstyle[aps,preprint,prb]{revtex}
\tighten
\begin{document}
\draft
\preprint{ITFA-94-24}
\title{Temperature relaxation and the Kapitza boundary resistance paradox}
\author{Alec Maassen van den Brink\cite{mail}}
\address{Institute for Theoretical Physics, University of Amsterdam,
Valckenierstraat 65,\\ 1018 XE Amsterdam, The Netherlands}
\author{H. Dekker}
\address{Institute for Theoretical Physics, University of Amsterdam,
Valckenierstraat 65,\\ 1018 XE Amsterdam, The Netherlands\\and\\TNO Physics
and Electronics Laboratory, P.O. Box 96864, 2509 JG Den Haag, The Netherlands}
\date{July 27, 1994}
\maketitle
\begin{abstract}
The calculation of the Kapitza boundary resistance between dissimilar
harmonic solids has since long (Little [Can.\ J. Phys.\ {\bf 37,} 334 (1959)])
suffered from a paradox: this resistance erroneously tends to a finite
value in the limit
of identical solids. We resolve this paradox by calculating temperature
differences in the final heat-transporting state, rather than with respect to
the initial state of local equilibrium. For a one-dimensional model
we thus derive an exact, paradox-free formula for the boundary resistance.
The analogy to ballistic electron transport is explained.
\end{abstract}
\pacs{05.60.+w, 44.30.+v, 67.40.Pm, 72.10.-d}

\section{Introduction}

The occurrence of a thermal resistance at an interface between two materials,
also known as Kapitza \cite{kapitza} resistance, is a well-known
phenomenon, in particular in low-temperature physics.
In fact, the present work has been prompted in the course of an investigation
of non-isothermal stochastic processes in Josephson junction devices.%
\cite{noniso}
The basic physical mechanism is partial
reflection of the phonons transporting the heat current. For the low
temperatures (in comparison with the Debye temperature) at which the boundary
contributes significantly to the total thermal resistance of a typical sample,
phonon wavelengths are much larger than the lattice spacing and a continuum
description of the materials should be adequate. In this case the reflection
process becomes a standard problem in elasticity theory, and using semi-%
classical arguments to determine phonon densities and fluxes the Kapitza
resistance was calculated along these lines in a classic paper by
Little.\cite{little}

For an ideal interface between the two elastic media and an infinitesimal
temperature difference between them so as to allow the use of linear
response theory, a fully quantum statistical calculation of the transport
process was given by Leung and Young.\cite{L&Y} Besides confirming the
intuitive
expectation that for this idealized system the semi-classical result should
be exact, the calculation is of interest in its own right as a non-perturbative
evaluation of a Kubo formula. In addition,
for a sample of finite size the dissipative
behaviour implicit in the term `resistance' was explicitly shown to hold only
up to times at which Poincar\'e recurrence sets in.

In Refs.\ \onlinecite{little} and \onlinecite{L&Y} the authors
noted their result for the Kapitza resistance to be
paradoxical in the limit of identical media: it tends to a finite
value instead of vanishing. They attributed this artifact
to the neglect of three-phonon scattering (which would take over as the
dominant origin of resistance in this case), arguing that without such
scattering the phonon mean free
path is infinite so that it would become problematic
to describe the system in terms
of local temperatures. This explanation is unsatisfactory for
the following reasons. {\em First of all}, three-phonon scattering is effective
as a cause of heat resistance and thermalization only when Umklapp processes
are
possible.\cite{kittel} Thus, one would have to endow the model with a lattice
structure as well. However, as mentioned above, this should not be relevant.
{\em Secondly}, the phonon mean free path is infinite for all values of the
energy transmission coefficient ${\cal T}$, and the authors
of Refs.\ \onlinecite{little}
and \onlinecite{L&Y} rightly do not suspect their results in the case of small
${\cal T}$. {\em Thirdly}, the
vanishing of the boundary resistance for a homogeneous medium
is a matter of principle, and should
not depend on the relative magnitude of other resistance contributions.

In this article we show that the aforementioned paradox is resolved in a
simple manner by calculating temperature differences in the final,
heat-transporting state of the system, and not with respect to
the initial state of local equilibrium.
For clarity of exposition we limit ourselves here to one space dimension.
A report on the technically more involved three-dimensional case is currently
in preparation.\cite{thesis}
In section \ref{semi} the corrected formula will be obtained by
a semi-classical consideration in the spirit of Ref.\ \onlinecite{little}.
In Sec.\ \ref{linres} this new result will be derived in the Kubo
formalism. Apart from solving the Kapitza resistance paradox (for which
Leung and Young's Kubo formula itself needs to be amended)
the work of Ref.~\onlinecite{L&Y} will be extended by giving the full
space and time dependence of the heat current. In the one-dimensional case the
relaxation to a stationary non-equilibrium state is thus shown to be
degenerate, namely jump-wise. Also, one will see that the semi-classical
result holds without the necessity of coarse-graining,
partly because the problem has no length scale apart from $\beta\hbar c$
(where $c$ is the sound velocity). The issue discussed in
this paper has a striking analogy\cite{buttiker,rolf92}
in ballistic electron transport.\cite{imry,stone,beenakker}
This analogy is the subject of Sec.~\ref{analogy}.
Some concluding remarks are made in Sec.\ \ref{remarks}.

\section{Semi-classical analysis of temperature relaxation}
\label{semi}

Consider two semi-infinite harmonic strings,\cite{L&Y,hans,L&Y2}
joined at $x=0$, with
sound velocities $c_{\rm l}$ ($c_{\rm r}$) for $x<0$ ($x>0$). For this
system the boundary conductance $K$ was defined in Ref.~\onlinecite{L&Y} as
$K\equiv{\cal J}/\Delta T$, where ${\cal J}$ is the energy
current in the {\em stationary\/} state which is reached after the system has
started at time $t=0$ in a state of local equilibrium with temperatures
$T_{\rm r}$ for $x>0$ and $T_{\rm l}=T_{\rm r}+\Delta T$ for $x<0$.
Their result (Eq.\ (17)) is $K=\frac{\pi}{6\hbar}{\cal T}T$, with ${\cal T}$
given in (\ref{calT}).
The ensuing value for $K$ in the case of identical strings (i.e.,
${\cal T}= 1$) is easily understood
using a semi-classical picture (Ref.~\onlinecite{little})
in which energy-carrying fields propagate according to the wave equation.
In view of the absence of an interface, excess energy at $x<0$ will then
be radiated half to $x=-\infty$, half to positive $x$. One thus arrives at
$K=\frac{1}{2}\,c\,C$, where $C$ is the heat capacitance per
unit length $C=\pi T/3\hbar c$. This indeed gives the finite value
$K=\frac{\pi}{6\hbar} T$, in contrast with the expected result $K=\infty$.
The finite heat current is a consequence of the finite specific heat of the
quantum string, combined with the finite speed of propagation (as opposed to
diffusive transport).

Now consider the general case of unequal
strings, with initial thermal energy density $\varepsilon_{\rm l}$
($\varepsilon_{\rm r}$)
($\equiv\langle{\cal H}_{\rm l,r}\rangle$, where $\langle{\cal H}\rangle$ is
the expectation value of the Hamiltonian density; see Sec.~\ref{linres})
for $x<0$ ($x>0$) at time $t=0$. At any $t=t_1>0$ the situation is as in Fig.\
\ref{string}.
Properly accounting for left- and right-going radiation, the expression for
the steady state value of the energy density $\varepsilon_i'$ ($i={\rm l,r}$)
is obtained by considering the history
of an energy packet $\varepsilon_i'\,\delta x_i$ and noticing that
$\delta x_{\rm r}/\delta x_{\rm l}=c_{\rm r}/c_{\rm l}$.
The result reads:
\begin{equation}
\varepsilon_{\rm l}'=\frac{\varepsilon_{\rm l} + {\cal R}\,
\varepsilon_{\rm l}+{\cal T} \varepsilon_{\rm r}
c_{\rm r}/c_{\rm l}}{2},\quad
\varepsilon_{\rm r}'=\frac{\varepsilon_{\rm r} + {\cal R}\,
\varepsilon_{\rm r}+{\cal T} \varepsilon_{\rm l}
c_{\rm l}/c_{\rm r}}{2}\;,
\end{equation}
with ${\cal R}=1-{\cal T}$.
Again putting $T_{\rm l} = T_{\rm r} + \Delta T$ ($\Delta T\ll T_i$),
$\varepsilon_i^{(\prime)}=\pi{T_i^{(\prime)}}^2/6\hbar c_i$
---where we have used the specific heat as given above---, one gets
\begin{equation}
\left.\begin{array}{l}
T_{\rm l}'=T_{\rm r}+(1-{\cal T}\!/2)\,\Delta T\\
T_{\rm r}'=T_{\rm r}+({\cal T}\!/2)\,\Delta T
\end{array}\right\}
\Rightarrow\Delta T'={\cal R}\,\Delta T\;.
\end{equation}
With the boundary conductance defined as the ratio of the stationary heat
flux and the {\em stationary\/} temperature jump, one now finds
\begin{equation}
K=\frac{{\cal J}}{\Delta T'}=
  \frac{{\cal J}}{\Delta T}\,\frac{\Delta T}{\Delta T'}=
  \frac{\pi}{6\hbar}T\frac{{\cal T}}{{\cal R}}\;,\label{ourresult}
\end{equation}
which indeed diverges for ${\cal T}\rightarrow 1$.

\section{Temperature relaxation in the Kubo formalism}
\label{linres}

The result (\ref{ourresult}) can also be calculated in the Kubo formalism.
That calculation (Sec.~\ref{drop}), however,
requires a nontrivial correction of the Kubo formula used in
Ref.\ \onlinecite{L&Y}.
In both Secs.~\ref{kuboform} and \ref{drop} we set $\hbar=1$.

\subsection{The Kubo formula}
\label{kuboform}

At $t=0$, to within first order in the temperature nonuniformity, the local
equilibrium density matrix is\cite{zubarev}
\begin{equation}
\rho_{\rm loc}=\rho_0\left\{1-\beta^{-1}\!\int_0^{\beta}\!\!d\tau
  \int\!d\mbox{\bf x}'\,\delta\beta(\mbox{\bf x}')
  \Bigl({\cal H}(\mbox{\bf x}',-i\tau)-
  {\langle{\cal H}(\mbox{\bf
x}',-i\tau)\rangle}_0\Bigr)\right\}\;,\label{rholoc}
\end{equation}
where $\beta^{-1}$ is the equilibrium temperature, $\rho_0=\exp(-\beta H)$ is
the equilibrium density matrix ($H$~is the full Hamiltonian),
${\cal H}$ is again the Hamiltonian density and ${\langle\cdot\rangle}_0$
indicates an equilibrium average. Note that (\ref{rholoc}) is not restricted
to the one-dimensional situation.
When calculating the energy current
${\langle{\cal J}(\mbox{\bf x},t)\rangle}_{\rm loc}$, the first and third
term of $\rho_{\rm loc}$ do not contribute, and one obtains
\begin{eqnarray}
{\langle{\cal J}(\mbox{\bf x},t)\rangle}_{\rm loc}&=
  &T^{-1}\!\int_0^{\beta}\!\!d\tau
  \int\!d\mbox{\bf x}'\,\delta T(\mbox{\bf x}')
  {\langle{\cal H}(\mbox{\bf x}',-i\tau)
  {\cal J}(\mbox{\bf x},t)\rangle}_0\label{step}\\
  &=&{\langle{\cal J}(\mbox{\bf x},0)\rangle}_{\rm loc}+
  \frac{i}{T}\int\!d\mbox{\bf x}'\,\delta T(\mbox{\bf x}')
  \int_0^t\!\!dt'\mathop{\rm Tr}\nolimits\left\{\rho_0\int_0^{\beta}
  \!\!d\tau\,e^{H\tau}
  [{\cal H}(\mbox{\bf x}'),H]e^{-H\tau}
  {\cal J}(\mbox{\bf x},t')\right\}\;.\nonumber
\end{eqnarray}
To begin with,
${\langle{\cal J}(\mbox{\bf x},0)\rangle}_{\rm loc}$ vanishes because of time
reversal invariance. In the second term, the trace can be taken as a single
correlation function; one would then get
${\langle{\cal J}(\mbox{\bf x},t)\rangle}_{\rm loc}\propto
\int_0^tdt'\,\langle\int_0^{\beta}\!d\tau{\cal J} (\mbox{\bf x}',-i\tau)
{{\cal J}(\mbox{\bf x},t')\rangle}_0$ without an additional $t'$
in the integrand.
However, in order to relate ${\langle{\cal J}\rangle}_{\rm loc}$ to
${\langle{\cal J}{\cal J}\rangle}_{\rm odd}$ we proceed differently and apply
the Kubo identity:
\begin{eqnarray}
{\langle{\cal J}(\mbox{\bf x},t)\rangle}_{\rm loc}&=
&\frac{i}{T}\int\!d\mbox{\bf x}'\,\delta T(\mbox{\bf x}')
  \int_0^t\!dt'\mathop{\rm Tr}\nolimits
  \left\{[\rho_0,{\cal H}(\mbox{\bf x}')]{\cal J}(\mbox{\bf x},t')
  \right\}\nonumber\\
  &=&\frac{i}{T}\int\!d\mbox{\bf x}'\,\delta T(\mbox{\bf x}')\int_0^t\!dt'
  \left\langle\left[{\cal H}(\mbox{\bf x}')+\int_0^{t'}\!dt''
  e^{-iHt''}\nabla{\cal J}(\mbox{\bf x}')e^{iHt''},
  {\cal J}(\mbox{\bf x})\right]\right\rangle_0\;,
\end{eqnarray}
where the step taken on the second line of Eq.\ (\ref{step}) has been repeated.
Integration by parts over $\mbox{\bf x}'$ and combining the
$t'$- and $t''$-integrals now yields
\begin{eqnarray}
{\langle{\cal J}(\mbox{\bf x},t)\rangle}_{\rm loc}&=
  &\frac{i}{T}\int\!d\mbox{\bf x}'\,\nabla T(\mbox{\bf x}')
  \int_0^t\!dt'\,(t-t')\,{\langle[{\cal J}(\mbox{\bf x},t'),
  {\cal J}(\mbox{\bf x}',0)]\rangle}_0 +\nonumber\\
  &&\hspace{5.5cm}\frac{it}{T}\int\!d\mbox{\bf x}'\,\delta T(\mbox{\bf x}')
  \left\langle\left[{\cal H}(\mbox{\bf x}'),
  {\cal J}(\mbox{\bf x})\right]\right\rangle_0\nonumber\\
  &=&\frac{2i}{T}\int\!d\mbox{\bf x}'\,\nabla T(\mbox{\bf x}')
  \int_0^t\!dt'\,(t-t')\,
  {\langle{\cal J}(\mbox{\bf x},t')
  {\cal J}(\mbox{\bf x}',0)\rangle}_{\rm odd} +\nonumber\\
  &&\hspace{5.5cm}\frac{it}{T}\int\!d\mbox{\bf x}'\,\delta T(\mbox{\bf x}')
  \left\langle\left[{\cal H}(\mbox{\bf x}'),{\cal J}(\mbox{\bf x})
  \right]\right\rangle_0\;.\label{kubo}
\end{eqnarray}
This Kubo formula will be crucial in Sec.\ \ref{drop}. Apart from the boundary
term, it differs from the one used in Ref.\ \onlinecite{L&Y} in that the
integrand presently involves $t'-t$ instead of $t'$.\cite{sign}

To be sure, this modification does not affect the value of the
stationary current. This follows from the explicit form for
${\langle{\cal JJ}\rangle}_{\rm odd}$ calculated in the Appendix:
\begin{equation}
{\langle{\cal J}(x,t){\cal J}(x',0)\rangle}_{\rm odd}=
\frac{{\cal T}}{i}\left\{
  \frac{1}{48\pi}\,\delta'''(t-\alpha)-\frac{\pi}{12\beta^2}\,
  \delta'(t-\alpha)\right\} +
  (\alpha\leftrightarrow-\alpha)\;,\label{jjodd}
\end{equation}
where $x>0$, $x'<0$, $\alpha\equiv x/c_{\rm r}-x'/c_{\rm l}$,
and where we returned to the one-dimensional case. Because the distributions
$\delta'''$ and $\delta'$ are odd functions of their arguments,
$\int_0^\infty dt'\,{\langle{\cal J}(t'){\cal J}\rangle}_{\rm odd}$
is seen to vanish.

In the boundary term of the Kubo formula (\ref{kubo}), the commutator
$[{\cal H},{\cal J}]$ will be localized at $\mbox{\bf x}'=\mbox{\bf x}$.
As will be shown explicitly in a simple case, this boundary term compensates
a second unphysical term arising from the lower limit of the $t'$-integration
in
the first part of Eq.~(\ref{kubo}). In this first part,
${\langle{\cal J}(\mbox{\bf x},t'){\cal J}(\mbox{\bf x}',0)\rangle}_{\rm odd}$
vanishes outside the `sound cone'. Thus, for $t'\downarrow 0$ only the
point $\mbox{\bf x}'=\mbox{\bf x}$ contributes in the $\mbox{\bf x}'$-%
integration. In the calculation of the energy drop in Sec.~\ref{drop},
$\delta T(x')$ vanishes identically near $x'=x>0$ and hence both unphysical
contributions to the current are absent. To show their cancellation for
arbitrary $\delta T(x')$ one needs the full space dependence of
${\langle{\cal JJ}\rangle}_{\rm odd}$. Therefore,
we will temporarily restrict ourselves to the case ${\cal T}= 1$,
where---by symmetry---formula (\ref{jjodd}) does hold for any $t$ and $\alpha$.
The equal-time commutator $\left\langle[{\cal H},{\cal J}]\right\rangle$
can be calculated using Wick's theorem or, equivalently, by
expanding the fields in creation and annihilation operators.\cite{anomaly}
The answer is:
\begin{equation}
  \left\langle\left[{\cal H}(x'),{\cal J}(x)\right]\right\rangle_0=
  \frac{-i}{c}\left\{\frac{\delta'''(\alpha)}{12\pi}-
                     \frac{\pi\delta'(\alpha)}{3\beta^2}\right\}\;.
\end{equation}
Inserting this in the second term of (\ref{kubo}), and evaluating the first
term of that formula to be
\begin{equation}
  \frac{1}{T}\int\!dx'\,\nabla T(x')\left\{\frac{\delta'(t-\alpha)}{24\pi}-
  \frac{\pi}{6\beta^2}\Bigl(\theta(t-\alpha)-\theta(-\alpha)\Bigr)-
  t\left(\frac{\delta''(\alpha)}{24\pi}-\frac{\pi\delta(\alpha)}
  {6\beta^2}\right)\right\} +
  (\alpha\leftrightarrow-\alpha)\;,\label{firstterm}
\end{equation}
the unphysical terms which grow linearly in $t$ are seen to cancel.

\subsection{The energy drop}
\label{drop}

For the case of arbitrary ${\cal T}$, the drop in energy
density $\Delta{\langle{\cal H}\rangle}_{\rm loc}$ can be calculated from
(\ref{kubo}) by using the detailed form of (\ref{jjodd}).
However, only a few general properties of
this correlation function will actually be needed (thereby suggesting the
validity of the present analysis
beyond the case of the 1D elastic continuum). Namely, by inspection of the
defining integrals it is easy to verify that for $t,\alpha>0$,
${\langle{\cal JJ}\rangle}_{\rm odd}$ is an odd function $f(s\equiv t-\alpha)$,
which is localized at $s=0$. As mentioned after Eq.~(\ref{jjodd}),
this immediately shows that the additional
$t$-term in the Kubo relation (\ref{kubo})
gives a vanishing contribution to the current, except at $t=\alpha$.\cite{sing}
On the other hand, omission of
this term would lead to a ${\Delta\langle{\cal H}\rangle}_{\rm loc}$
which is a factor two too large compared to the correct value obtained below in
Eq.\ (\ref{deltah}).

Let us now finally give a microscopic calculation of the
change in energy density. Again choosing $x>0$, one has
\begin{eqnarray}
\Delta{\langle{\cal H}(x)\rangle}_{\rm loc}&=&\int_0^\infty\!\!dt\,
  \partial_t{\langle{\cal H}(x,t)\rangle}_{\rm loc}\nonumber\\
  &=&-\frac{1}{c_{\rm r}}\int_0^\infty\!\!dt\,
  \partial_\alpha{\langle{\cal J}(\alpha,t)\rangle}_{\rm loc}\nonumber\\
  &=&\lim_{\tau\rightarrow\infty}\frac{{\rm Const}}{c_{\rm r}}
  \int_0^\tau\!dt\int_0^t\!dt'
  \,(t-t')\,\partial_\alpha f(t'-\alpha)\nonumber\\
  &=&\lim_{\tau\rightarrow\infty}\frac{{\rm Const}}{c_{\rm r}}\int_0^\tau\!dt
  \left[tf(-\alpha)-\int_0^t\!dt'f(t'-\alpha)\right]\nonumber\\
  &=&\lim_{\tau\rightarrow\infty}-\frac{{\rm Const}}{c_{\rm r}}\int_0^\tau\!dt
  \,(\tau-t)f(t-\alpha)\nonumber\\
  &=&\frac{1}{c_{\rm r}}{\langle{\cal J}(x,t=\infty)\rangle}_{\rm loc}\;.
\label{deltah}
\end{eqnarray}
In the fourth and fifth line the first term vanishes because of the properties
of $f$ mentioned above. With ${\left\langle{\cal J}\right\rangle}_{\rm loc}=
\frac{\pi}{6}{\cal T}\Delta T$, the value of
$\Delta{\langle{\cal H}\rangle}_{\rm loc}$ according to Eq.~(\ref{deltah})
coincides with the one obtained in the semi-classical analysis of
Sec.~\ref{semi}, leading to our result (\ref{ourresult}).

\section{Analogy with ballistic electron transport}
\label{analogy}

The analysis of electric current in a one-dimensional conductor
as a scattering problem was pioneered by Landauer.\cite{landauer}
Consider two well-separated reservoirs at electrochemical
potentials $\mu_{\rm l}=\varepsilon_{\rm F}+eV$ and
$\mu_{\rm r}=\varepsilon_{\rm F}$,
respectively. Let the reservoirs be connected by a one-%
dimensional lead, interrupted by a barrier (at $x=0$, say) with
transmission coefficient
${\cal T}$. Using the 1D density of states $1/hv_{\rm F}$ ($v_{\rm F}$
being the Fermi velocity),
at zero temperature the current per spin direction is obtained as
$I={\cal T}(ev_{\rm F})(eV/hv_{\rm F})=\frac{e^2}{h}{\cal T}V$.
This leads to a conductance $G=\frac{e^2}{h}{\cal T}$ in the case of a
voltage measurement at the reservoirs.

Near the barrier, however, the conductor
is not well characterized by the reservoir potentials $\mu_{\rm l,r}$.
For energies between
$\mu_{\rm l}$ and $\mu_{\rm r}$, at a position $x<0$ all left-going states are
occupied, but only a fraction ${\cal R}$ of the right-going
states is filled. This leads to an effective electrochemical potential
$\mu_{\rm l}'=\varepsilon_{\rm F}+eV(1+{\cal R})/2$. Similarly, for $x>0$
one gets $\mu_{\rm r}'=\varepsilon_{\rm F}+eV{\cal T}/2$.
Calculating the conductivity using $(\mu_{\rm l}'-\mu_{\rm r}')/e={\cal R}V$ as
the relevant voltage difference for a measurement near the barrier,
one finds the Landauer formula
\begin{equation}
G=\frac{e^2}{h}\frac{{\cal T}}{{\cal R}}\;.\label{landform}
\end{equation}

If one measures the voltage difference between the reservoirs,
the finite upper limit of
$G=\frac{e^2}{h}{\cal T}$ as ${\cal T}\uparrow 1$ makes perfect sense
physically. The resulting $G^{-1}=\frac{h}{e^2}$ has the significance of a
purely geometric contact resistance between the reservoirs. On the other
hand, $G$ in Eq.~(\ref{landform}) is the quantity which
describes the intrinsic properties of the barrier.

The analogy to the heat transport problem is now easily seen: the role of the
electric current is played by the heat current, the
electron densities from which the electrochemical potentials are
determined are the analogue of the energy densities considered in Secs.~%
\ref{semi} and \ref{linres}, and the effective
electrochemical potentials translate into local temperatures.

The appropriate definition of $\mu_i'$ is not trivial: one could argue that
local chemical potentials are only defined up to the difference in
potential pertaining to left- and right-going states
separately. This leads to a conceptual uncertainty
$(\Delta G)/G\approx 2{\cal T}$, and would therefore
allow a thermodynamic description of the current transport only if
${\cal T}\ll 1$. In that case the distinction between the two formulas for $G$
obviously loses its meaning. The analogous caveat holds for heat transport.%
\cite{bulk}
However, the potentials $\mu_{\rm l,r}'$ leading to Eq.~(\ref{landform})
are defined in such a way that, if the system would have been in local
equilibrium at these potentials, the electron density would be the same as
the actual density. This guarantees that the conductance (\ref{landform})
trivially satisfies the Einstein relation $G=e^2\partial_\mu n_{\rm p} D$,
where the diffusivity $D$ relates the particle current ${\cal J}_{\rm p}$
to the particle density jump across the barrier $\Delta n_{\rm p}$
by ${\cal J}_{\rm p}=D\Delta n_{\rm p}$.\cite{imry}
Moreover, this definition is the correct one in the following sense:
if one couples
two additional reservoirs infinitesimally weakly to the left and right of the
barrier, the condition that no current will flow
is that they must be at the potential $\mu_{\rm l}'$ and $\mu_{\rm r}'$,
respectively (see Ref.\ \onlinecite{imry} for the precise statement). In our
case one could---at least in principle---probe local temperatures by
weakly coupling additional heat baths in a similar way.

In the system described in Secs.\ \ref{semi} and \ref{linres}, the role of the
reservoirs is played by the infinite leads themselves. For electron transport,
this way of modeling the reservoirs was introduced earlier in
Ref.\ \onlinecite{stone}.
The criterion for a good reservoir is that it should only re-emit electrons
(phonons) after complete thermalization and loss of phase memory.
The leads satisfy this condition by never re-emitting any particle.
Violation of this condition, for example by taking finite leads, will
lead to different results, as mentioned already in the Introduction.
%

The absolute temperature, which
enters as a prefactor in our formula for the boundary conductance,
has no counterpart in the current transport problem.
Its presence is a consequence of the Bose statistics obeyed by the phonons
(see also the denominator in (\ref{jjint})), as opposed to the Fermi statistics
of electrons.

\section{Concluding remarks}
\label{remarks}

The reasoning leading to the paradox-free formula (Eq.\ (\ref{ourresult}))
for the Kapitza boundary conductance $K$
calculated in this article is not invalidated by the presence of a
thermalization mechanism like the weak three-phonon scattering contemplated in
Refs.\ \onlinecite{little} and \onlinecite{L&Y}.
On the contrary, such a mechanism rather brings the local energy
distribution closer to a thermal one (under the condition that there should be
a heat current~${\cal J}$), and thus only strengthens the case for
$\varepsilon_i'$ as the energy from which stationary temperatures should be
calculated.

{\it In conclusion\/}: using the model of coupled harmonic strings,
we have solved the Kapitza boundary resistance paradox
(Little's formula) by means of an exact calculation of the properties of the
interface in the final steady state, leading to the new
formula Eq.\ (\ref{ourresult}).
Besides, we have explained the analogy between local temperature measurements
in the heat transport problem and voltage measurements in ballistic electron
transport (Landauer's formula).

\section*{Acknowledgements}

We thank P. T. Leung and K. Young (Hong Kong), and R. Landauer (IBM,
Yorktown Heights) for their interest in this work.

\appendix
\section*{The current-current correlation function}

We calculate (\ref{jjodd}) along the lines of Ref.\ \onlinecite{L&Y}.
With the Hamiltonian density
${\cal H}=\frac{1}{2\rho}\pi^2+\frac{M}{2}(\nabla\phi)^2$ (where $\phi$ is the
field operator, $\pi$ is its conjugate momentum, $\rho$ is the mass density,
and $M$ is the modulus of elasticity), local energy conservation implies
${\cal J}=-\frac{M}{2\rho}\left\{\pi,\nabla\phi\right\}$ for the
energy current.\cite{herm} Using this expression for ${\cal J}$,
${\langle{\cal JJ}\rangle}_{\rm odd}$ can be expanded in
terms of field-field correlation functions using Wick's theorem.
By a KMS-relation, the field-field correlation functions can be obtained from
the field-field commutators only, and these in turn follow from
the classical wave equation. Carrying out this procedure leads to
\begin{equation}
  {\langle{\cal J}(x,t){\cal J}(x',0)\rangle}_{\rm odd}=\frac{{\cal T}}{2i}
  \int\!\frac{d\omega}{2\pi}\int\!\frac{d\omega'}{2\pi}\,
  \frac{\omega\omega'\sin\Bigl[(\omega+\omega')(t-\alpha)\Bigr]}
       {(1-e^{-\beta\omega})(1-e^{-\beta\omega'})}\;+\;
  (\alpha\leftrightarrow-\alpha)\;,\label{jjint}
\end{equation}
where ${\cal T}$ is defined in terms of the impedances $Z_i\equiv\rho_ic_i$ as
\begin{equation}
{\cal T}=\frac{4Z_{\rm l}Z_{\rm r}}{(Z_{\rm l}+Z_{\rm r})^2}\;,\label{calT}
\end{equation}
and $x>0$, $x'<0$ as in the main text. The infrared divergence
present in the field-field correlation functions because the energy of an
oscillation vanishes in the long-wavelength limit, is seen to be absent
in the current-current correlation function as such long-wavelength
excitations hardly contribute to the energy transport. On the other hand, the
ultraviolet divergence caused by taking the continuum limit for the string
must be dealt with, e.g., by regularizing
${\langle{\cal JJ}\rangle}_{\rm odd}$ using a
spatial smoothing. Taking a Lorentzian for that purpose leads to a cutoff
$e^{-\eta|\omega+\omega'|}$ in the integral in (\ref{jjint}). However, since
the integrand vanishes as $\sim e^{-\beta|\omega^{(\prime)}|}$ in the half
plane
$\omega+\omega'<0$ one may as well use the cutoff $e^{-\eta(\omega+\omega')}$.
The integral then factorizes and the ensuing single integrals can be done
exactly. With $\beta=1$, this yields
\begin{eqnarray}
  \int\!\frac{d\omega}{2\pi}\int\!\frac{d\omega'}{2\pi}\,
  \frac{\omega\omega'\sin\Bigl[(\omega+\omega')t
  \Bigr]e^{-\eta(\omega+\omega')}}
       {(1-e^{-\omega})(1-e^{-\omega'})}&=&
  \mathop{\rm Im}\nolimits
  \left[\frac{\pi}{2\sin^2(\pi(\eta-it))}\right]^2\nonumber\\
  &=&\left\{\frac{1}{6}\partial_t-\frac{1}{24\pi^2}\partial_t^3\right\}
    \mathop{\rm Im}\nolimits\frac{1}{t+i\eta}+
    \mathop{\rm Im}\nolimits
    \left[{\cal O}\left((t+i\eta)^0\right)\right]\nonumber\\
  &=&\frac{1}{24\pi}\delta'''(t)-\frac{\pi}{6}\delta(t)
  \quad\text{for }\eta\downarrow 0\;.
\end{eqnarray}
Scaling back to $\beta\neq 1$ and inserting this result into (\ref{jjint}),
one obtains ${\langle{\cal JJ}\rangle}_{\rm odd}$ as given in Eq.\
(\ref{jjodd}).

\begin{figure}
\caption{Two harmonic strings (with an interface at $x=0$) at a time $t=t_1$
after the system started in a different local equilibrium on either
side of $x=0$.}
\label{string}
\end{figure}


\begin{references}
\bibitem{sign} In the fourth line below Eq.\ (2) of Ref.\
  \onlinecite{L&Y}, the labels $z>z_1$ and $z<z_1$ should be
  interchanged. Apart from this, all signs are consistent.
\bibitem{anomaly} A calculation like
  $\left[{\cal H}(x'),{\cal J}(x)\right]=
  \left[\frac{\pi(x')^2}{2\rho}+\frac{M}{2}(\partial_{x'}\phi(x'))^2,
  -\frac{M}{2\rho}\left\{\pi(x),\partial_x\phi(x)\right\}\right]=$
  {\Big(}by repeated use of $\left[\phi(x),\pi(x')\right]=
  i\,\delta(x-x')${\Big)}
  $=i\left\{\frac{M}{\rho^2}\pi(x')\pi(x)+
  \frac{M^2}{\rho}\,\partial_{x'}\phi(x')
  \,\partial_{x}\phi(x)\right\}\delta'(x-x')=
  -2ic^2\,\delta'(x'-x)\,{\cal H}(x)+
  ic^2\,\delta(x'-x)\,\partial_x{\cal H}(x)\Rightarrow\left\langle
  \left[{\cal H}(x'),{\cal J}(x)\right]\right\rangle=
  \frac{i\pi}{3c\beta^2}\delta'(\alpha)$
  is {\em not\/} correct, because the Hamiltonian densities arising at the end
  of this exercise are not normally ordered
  (even if one would start with ${:}{\cal H}{:}$, since
  ${\cal H}-{:}{\cal H}{:}\propto\openone$,
  which of course commutes with ${\cal J}$).
  This means that in the above formula one is multiplying delta functions
  with an infinite object, which can only be done reliably by means of the more
  elaborate method of the main text. The additional, temperature-independent
  contribution which is found there (and the analogous one for
  ${\langle{\cal JJ}\rangle}_{\rm odd}$ in Eq.\ (\ref{jjodd})) is
  known as a Schwinger term. See, e.g., C. Itzykson,
  J.-B. Zuber, {\it Quantum Field Theory\/} (McGraw-Hill, Singapore, 1988).
\bibitem{sing} Because the prefactor $t-t'$ multiplying
  ${\langle{\cal JJ}\rangle}_{\rm odd}$ in Eq.~(\ref{kubo}) has a first order
  zero at $t'=t=\alpha$, ${\langle{\cal J}\rangle}_{\rm loc}$ as obtained
  from Eq.~(\ref{kubo}) is less singular than it would have been without the
  $t$-term: it contains $\delta'(t-\alpha)$ (see also Eq.~(\ref{firstterm}))
  but not $\delta''(t-\alpha)$.
\bibitem{bulk} This subtlety is absent in the study of bulk transport, where
one can unambiguously define local quantities by averaging over a spatial
region
which is much larger than a thermalization length. See also the Introduction.
\bibitem{herm} Notice that the unsymmetrized Eq.~(7) from Ref.~\onlinecite{L&Y}
  would incorrectly imply that ${\cal J}$ is not hermitian.
\end{references}
\end{document}